\documentclass[aps,pre,amsmath,amssymb,reprint,floatfix]{revtex4-1}
\usepackage{lipsum}
\usepackage{amssymb,euscript,latexsym,amsmath,amsthm}
\usepackage{graphicx}
\usepackage{color}
\usepackage{setspace}
\usepackage{float}
\usepackage{bm}
\usepackage[dvipsnames]{xcolor}
\usepackage{chngpage}
\usepackage{hyperref}
\usepackage{subcaption}
\usepackage{cancel}
\usepackage{setspace}
\usepackage{graphicx}
\usepackage[T1]{fontenc}
\usepackage[utf8]{inputenc}
\usepackage[font=small,labelfont=bf,tableposition=top]{caption}
\usepackage{booktabs,multirow}
\usepackage{gensymb}
\usepackage{natbib}
\usepackage{mathtools}

\DeclareGraphicsExtensions{.png,.ps,.pdf}

\def\*{\cdot}

\begin{document}

\title{Sea star inspired crawling and bouncing}

\author{Sina Heydari$^1$}
\author{Amy Johnson$^2$}
\author{Olaf Ellers$^2$}
\author{Matthew J. McHenry$^3$}
\author{Eva Kanso$^1$}
\email{kanso@usc.edu}
\affiliation{$^1$Department of Aerospace and Mechanical Engineering,  \\ University of Southern California, 854 Downey way, Los Angeles, California 90089, USA\\
$^2$Department of Biology, \\Bowdoin College, 255 Maine Street, Brunswick, Maine 04011, USA \\
$^3$Department of Ecology and Evolutionary Biology, \\ University of California, Irvine, 321 Steinhaus Hall, Irvine, California 92697, USA}

\date{\today}
\begin{abstract}
The oral surface of sea stars is lined with arrays of tube feet that enable them to achieve highly controlled locomotion on various terrains.
The activity of the tube feet is orchestrated by a nervous system that is distributed throughout the body without a central brain. How such a distributed nervous system produces a coordinated locomotion is yet to be understood. We develop mathematical models of the biomechanics of the tube feet and the sea star body. In the model, the feet are coupled mechanically through their structural connection to a rigid body. We formulate hierarchical control laws that capture salient features of the sea star nervous system. Namely, at the tube foot level, the power and recovery strokes follow a state-dependent feedback controller. At the system level, a directionality command is communicated through the nervous system to all tube feet. We study the locomotion gaits afforded by this hierarchical control model. We find that these minimally-coupled tube feet coordinate to generate robust forward locomotion, reminiscent of the crawling motion of sea stars, on various terrains and for heterogeneous tube feet parameters and initial conditions. Our model also predicts a transition from crawling to bouncing consistent with recent experiments. We conclude by commenting on the implications of these findings for understanding the neuromechanics of sea stars and their potential application to autonomous robotic systems.
\end{abstract}

\maketitle

\section{Introduction}
\label{sec:intro}

\begin{figure*}[!t]
\centering
\includegraphics[width = \linewidth]{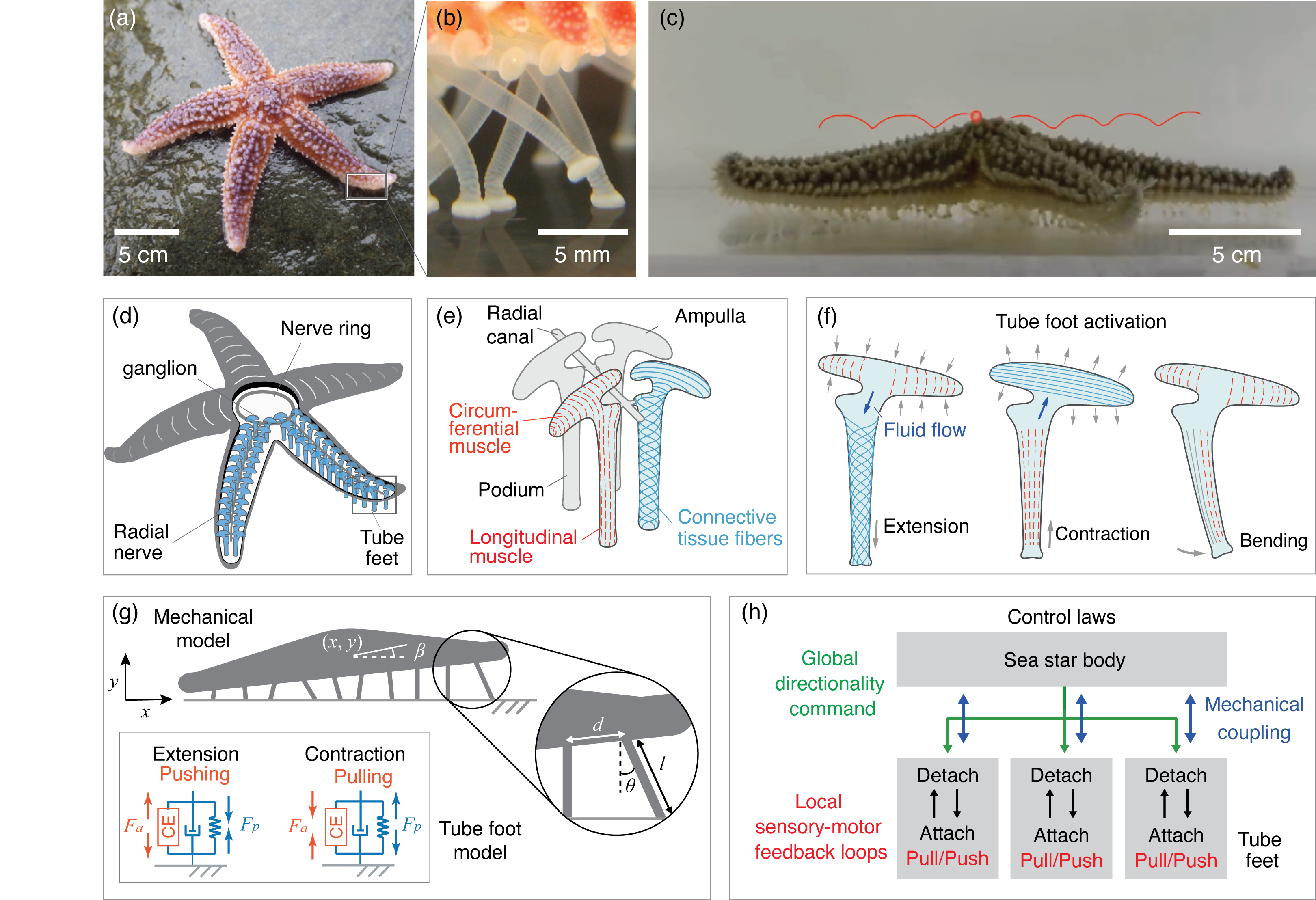}
\caption{\footnotesize \textbf{Sea stars:} { (a)  The common sea star \textit{Asterias rubens} (source: Shutterstock), (b) close-up on the tube feet lining the ventral surface of \textit{Asterias rubens} (source: Symbiotic service, San Diego), (c) bounce gait in \textit{Asterias forbesi}~\cite{Pennisi2019}, (d) schematic of \textit{Asterias rubens}, showing nervous system comprising a circumoral nerve ring and radial nerves, (e) tube foot anatomy for an adult sea star, (f) muscles are innervated by neurons located in the radial nerves and nerve ring. Activation of podia and ampulla muscles lead to contraction, extension and bending of the tube feet, (g) schematic of our mechanical model of the sea star and tube feet inspired actuators, with inset showing contractile, passive and dissipative force elements along each tube foot, (h) hierarchical motor control of the tube feet consisting of {\em global} directionality commands issued by the radial nerves and nerve ring and {\em local} sensory-motor feedback loops at the tube foot level. } } 
\label{fig:seastar}
\end{figure*}

Echinoderms are a group of marine invertebrates that use tube feet to achieve remarkable locomotion tasks. 
Sea stars, for example, have an oral surface that is lined with hundreds of tube feet used to crawl on various terrains, from smooth sand and glass surfaces to rocky substrates, see Fig.\ref{fig:seastar}. To achieve these feats of locomotion, individual tube feet are equipped with integrated sensing and actuation, and the activity of arrays of tube feet is orchestrated by a nervous system that is distributed throughout the body. {How the distributed nervous system and numerous tube feet interact to give rise to coordinated motion has long been a question of interest for researchers. In 1945, Smith put forward a plan of neuron configuration and axon distribution based on behavioral experiments and neuroanatomy~\cite{Smith1945}. 
Lacking a brain, the central nervous system comprises a ring nerve at the center of the body with radial nerves that innervate the tube feet and extend to a simple eye at the distal tips of each arm and innervates the tube feet~\cite{Bullock1965, Cobb1987b, Mashanov2016, Zueva2018}.} 
The behavior of tube feet was studied later by recording the stepping phases -- power and recovery strokes -- that each tube foot undergoes during locomotion \cite{Kerkut1953, Kerkut1954, Kerkut1955, Smith1947}. While all tube feet step in the same direction during walking, Kerkut's studies showed an absence of determinate phase relationship in the steps of different feet, suggesting the ability for individual action within each tube foot \cite{Paine1929, Bullock1965}. Taken together, these experimental findings hint at the presence of a hierarchical structure within the nervous system of sea stars. There seems to be a central communication from the radial and ring nerves through which a dominant direction of motion emerges, while the tube feet are individually capable of sensing and actuation.

More recently, there has been a growing effort to understand distributed control in biology,  in part due to their potential applications in autonomous robotic systems \cite{Mao2014, Kano2017, Owaki2017, Levy2015, Yasui2017}. Specifically, there have been multiple studies on how a direction of motion emerges from the distributed nervous systems in echinoderms, such as brittle stars and sea urchins \cite{Astley2012, Matsuzaka2017, Yoshimura2018, Clark2018, Zueva2018, Kano2019}. These studies, although acknowledge the hypothesis of a hierarchical control mechanism in echinoderms,  focus mostly on the centralized, system-level control, namely the directionality command and how it is transferred through the nerve ring. They lack details on how localized sensing and actuation at the tube feet level comes into play. 

In this study, we  introduce a mathematical model of sea star locomotion based on hierarchical control laws with local sensory-motor feedback loops at the tube foot level and a global directionality command at the system level. 
These control laws are implemented in mechanical models of the sea star that take into account salient features of the tube feet biomechanics as muscular hydrostats with no rigid skeletal support~\cite{Kier1992}. Each tube foot is modeled as a soft actuator that generates state-dependent active forces.  The tube feet control has no explicit communication of state between tube feet. Each tube foot is an autonomous entity that receives a global command about the direction of motion. Besides a shared directionality command, the tube feet are coupled only structurally through their attachment to a rigid body representation of the sea star.

We examine the sea star locomotion in the context of this mathematical model.
We particularly focus on two distinct modes of locomotion exhibited by sea stars: crawling and bouncing.  When stimulated, sea stars across various species are reported to exhibit a bounce gait in which they coordinate their feet to increase their speed~\cite{Johnson2019,Ellers2014,Ellers2018,Etzel2019}.  This bounce gait is characterized by amplified vertical oscillations and a discernible frequency and wavelength of motion; see Fig.~\ref{fig:seastar}(c). On the other hand, the crawl gait has a lower locomotion speed, dampened oscillations, and irregular trajectory of motion for which it is difficult to identify a frequency and wavelength. 
The bounce gait, which usually happens when tens of tube feet synchronize into three groups, raises new and interesting questions. Is there an underlying mechanism for sea stars to coordinate not only their direction of motion, but also the actuation of tens of tube feet? Or does the transition to bouncing happen as a result of the collective dynamics of individual and minimally-coupled tube feet? We address these questions by performing numerical experiments based on our mathematical model.

\begin{table}[!b]
\caption{Sea star parameters (based on~\cite{Smith1946, Kerkut1953, Appelhans2012})
}
\begin{tabular}{c|ccc}
\hline
 &  adult   &   \\
 & \textit{Asterias rubens} & \\
\hline
body diameter &  10--30 cm&  \\
wet weight &  3.25--6 g&  \\
dry weight &  9--15 g&  \\
Number of tube feet & $\approx$ 1000&  \\
Tube feet length
& 1.25--8 mm &  \\ 
force per tube foot & weight/0.1$\times$(number of tube feet)&\\
\hline
\end{tabular}
\label{table:params}
\end{table}

The organization of this work is as follows. In~\S\ref{sec:model}, we develop an abstract representation of the tube feet as soft actuators that can generate active pushing and pulling forces, and we model the sea star as a rigid body connected to an array of soft actuators. Though the model is abstract, we choose parameter values consistent with measurements of the common sea star \textit{Asterias rubens} given in table \ref{table:params}. An adult \textit{Asterias rubens} usually grows up to be $10$--$30$ cm in diameter, with five arms each equipped with hundreds of tube feet. Specifically, we choose the parameter governing the active force per tube foot to be consistent with Kerkut estimation that only $10 \%$ of the total number of the tube feet are needed to support the sea star's submerged weight \cite{Kerkut1953}. 
 We mathematically couple the hierarchical control laws described above to the equations of motion governing the body mechanics. 
The results of the models are presented and discussed in~\S\ref{sec:results}. We conclude in~\S\ref{sec:conc} by commenting on the advantages and limitations of our modeling approach and on the implications of our findings for understanding {the distributed nervous systems of echinoderms} and for developing soft robotic systems.

\section{Mathematical Modeling}
\label{sec:model}

\subsection{Tube Feet Mechanics}
\label{sec:tubefeet}

Each tube foot consists of a cylindrical channel, called a podium, capped by a bladder-like structure, called an ampulla; see Fig.~\ref{fig:seastar}(d-f).
The interior space of the ampulla is continuous with the interior of the podium, such that interstitial fluid moves freely between these two spaces. The walls of both the podium and ampulla includes layers of connective-tissue fibers that are stiff in tension (light blue lines in Fig.~\ref{fig:seastar}e-f), and superficial layers of muscle that serve to generate tension in the direction of the muscle fibers (orange lines in Fig.~\ref{fig:seastar}e-f).
In the podium, the connective-tissue fibers are arranged helically to favor elongation of the podium under pressure, and the muscle fibers are arranged longitudinally \citep{Mccurley1995}. The ampulla is characterized by longitudinally-oriented connective-tissue fibers and circumferential muscles.

Experimental observations suggest that the podium is extended by contraction of the circumferential muscles in the ampulla. This action generates pressure that expels the interstitial fluid from the ampulla into the podium (Fig.~\ref{fig:seastar}f).  Relaxation of the ampullar muscles causes the podium to retract. Retraction of the podium can be continued further through active contraction of the podium's longitudinal muscles, which expels water from the podium into the ampulla.  Further, a subset of these muscles could be activated to presumably bend the podium, provided the circumferential muscles of the ampulla maintain tension to prevent fluid flow from the podium. 
This model for the biomechanics of individual tube feet provides a starting-point for a mathematical description of these biological soft actuators and the premise for designing engineered counterparts.

It is worth noting that the principles of operation of the tube feet as muscular hydrostats share similarities with  pneumatic artificial muscles such as the McKibben actuators that convert hydraulic pressure into mechanical work. 
A mathematical relationship between the tensile forces and the length of these actuators can be obtained from first-principles~\cite{Chou1996, Tondu2012}. Similarly, force generation in the tube feet can be modeled by taking into account the balance between fluid pressure and wall stress in the ampulla and podium~\cite{McHenryinprep, Mccurley1995}. Our goal here is to formulate an abstract model of each tube foot as an actuator capable of producing active pushing and pulling forces, without looking into the details of force generation by muscle activation in the ampulla and podium.

To mathematically describe the behavior of a tube foot, we must model the forces it generates during its power and recovery stroke, that is, we must model its attachment and detachment dynamics. We postpone the attachment-detachment issue to \S\ref{sec:control}.  
To fix ideas, we consider a weight-carrying tube foot with the base of the podium attached to a flat horizontal plane. We assume that the tube foot cannot bend actively when attached; in other words, it cannot generate active moments during attachment, only active longitudinal forces. By contracting the ampulla and extending the podium, the tube foot produces \textit{an active pushing force}; more precisely, by the law of action and reaction, the tube foot produces a pair of forces pushing onto both the  plane of attachment and the load it is carrying. Inversely, an \textit{active pulling force} can be generated by contracting the podium and expanding the ampulla. Clearly, active pulling requires additional contact forces to ensure the podium maintains contact with the ground, through friction, suction, or chemical adhesion~\citep{Hennebert2012, Nachtigall2013, Lengerer2019}. This active force model can be thought of as a state-dependent controller, where the magnitude and sign of the active force depends on the state of the tube foot, namely, its length and activation mode (pushing or pulling), while its direction is always acting longitudinally along the tube foot.
In tandem with these active pushing and pulling forces, the tube foot experiences restoring elastic forces due to the connective tissues. Its extension or contraction is dampened by viscous resistance due to the interstitial fluid movement. Put together, each tube foot can be modeled as a soft actuator with (i) an active force generating element $F_a$ that is either {\em pushing} or {\em pulling}, (ii) a passive restoring force element $F_p$, and (iii) a viscous damping element $F_d$, all acting along the length of the tube foot, as shown in the inset of Fig.~\ref{fig:seastar}(g).

\begin{figure}[!t]
\centering
\includegraphics[scale = 1]{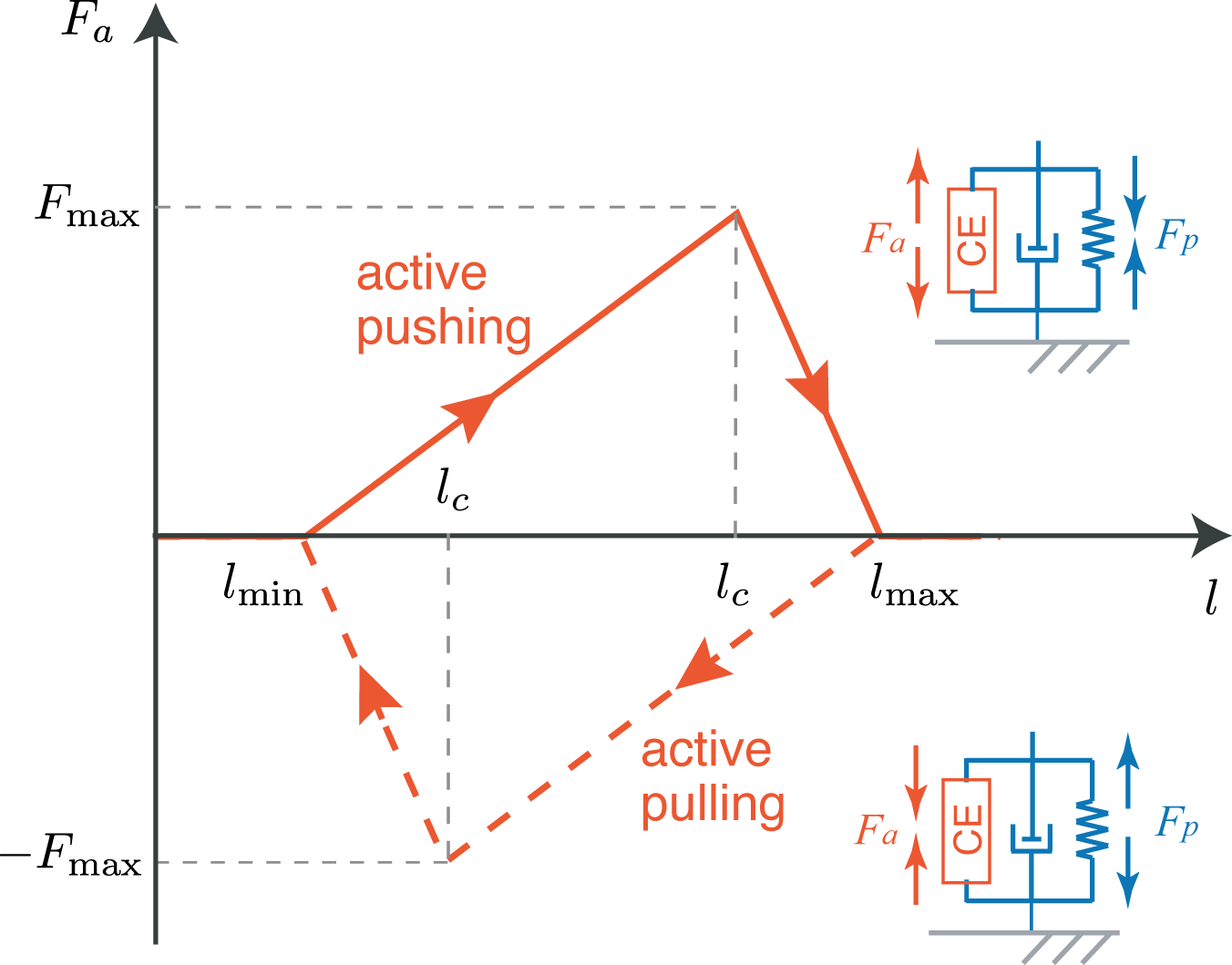}
\caption{\footnotesize {\footnotesize \textbf{Tube-foot inspired soft actuator:} when attached to a substrate, a tube foot generates either a pushing or a pulling force on the body it is attached to. These active forces act longitudinally along the tube foot direction and their magnitude depends on the tube foot length $l$.
 } } 
\label{fig:Fa}
\end{figure}

Let $l$ be the length of the tube foot, with $l_{\rm min}$ and $l_{\rm max}$ being its minimum and maximum length. We consider the restoring elastic force $F_p$ to be linear $F_p = -k_p(l-l_o)$, where $l_o$ is the length at which the connective fibers are un-stretched. We also consider a linear damping force of the form $F_d = -c_d \dot{l}$. 
Inspired by Hill's muscle model~\cite{Hill1938,Fung1982}, we use a piecewise linear force-length relation to model the active force $F_a$ generated in the tube foot, namely, we write 
\begin{equation}
 F_a = F_{\rm max} \Phi(l)
\label{eq:fa}
\end{equation}
where $F_{\rm max}$ is a scalar constant denoting the maximum force generated in the tube foot, and $\Phi(l)$ is a length-dependent function that describes the force profile. We let $l_c$ denote the length at which the active force is maximum as shown in Fig.~\ref{fig:Fa}. When in a pushing state, $\Phi(l)$ is given by
\begin{equation}
\begin{split}
\Phi_{\rm push}(l) = \begin{cases} 
 \dfrac{(l - l_{\rm min})}{(l_c - l_{\rm min})},  \ &  l_{\rm min} <l<l_c, \\[3ex]
 \dfrac{(l - l_{\rm max})}{(l_c - l_{\rm max})},  \ & l_c<l< l_{\rm max}, \\[3ex]
 \quad 0,   & l< l_{\rm min}$ \ \textrm{and} \  $l> l_{\rm max}.
\end{cases} 
\end{split}
\label{eq:phi}
\end{equation}
Similar expressions can be obtained for pulling; the pushing and pulling force profiles are shown in Fig.~\ref{fig:Fa} as a function of length. Here, the pushing and pulling force profiles are symmetric.

Sea stars employ tube feet to generate a diverse array of motion. However, it is instructive first to explore the theoretical situation of vertical extension and contraction of a single tube foot carrying a weight $mg$, where $m$ is mass and $g$ is the gravitational constant.  In this vertical ``standing'' regime, the length of the tube foot $l$ coincides with the vertical position $y$ of the mass. The equation of motion can be obtained from a straightforward application of Newton's third law
\begin{equation}
F_a - k_p(l-l_o) -c_d \dot{l} - \alpha mg = m \ddot{l}.
\label{eq:vertical}
\end{equation}
Here, we introduced a parameter $\alpha = (1-{\rho}/{\rho_s})$ to account for the buoyancy effects by considering the densities $\rho$ and $\rho_s$ of water and the sea star, respectively, with $\rho/\rho_s < 1$. The parameter $\alpha = (1-{\rho}/{\rho_s}) \in [0,1]$: $\alpha = 1$ corresponds to the dry weight of the sea star and $\alpha =0$ corresponds to a neutrally-buoyant sea star. Without loss of generality,  we set $\alpha =1$ while the value of $mg$ can be set independently.

It is useful for writing the equations of motion in non-dimensional form to introduce the length scale $L = l_{\textrm{max}} - l_{\textrm{min}}$.  
We also introduce two time scales: an inertial time scale $T_g = \sqrt{L/g}$ obtained by balancing the weight and inertial forces ($mg \sim mL/T_g^2$) and a relaxation time scale $T_d=c_d/k_p$ obtained by balancing the damping and passive spring forces ($c_d L/T_d \sim k_p L$).
Small values of $T_g$ describe a system where the weight is large compared to the inertial forces, whereas large values of $T_d$ imply that damping is dominant. Observations of sea star locomotion suggest strong damping and weak inertial forces. We thus choose $T_g <1$ and $T_d >1$ such that the non-dimensional ratio $ T_d/T_g$ is larger than 1.

We rewrite Eq.~\eqref{eq:vertical} in non-dimensional form using the length scale $L = l_{\textrm{max}} - l_{\textrm{min}}$, and the relaxation time scale $T_d = c_d / k_p$,
\begin{equation}
\mu \ddot{{l}} + c_d \dot{{l}} + k_p ({l} - {l}_o)  = {F_a} - {mg} .
\label{eq:nondim_vertical}
\end{equation}
Here, all parameters and variables are non-dimensional. Specifically, $c_d =1$, $k_p =1$, and ${F_a}$ and $mg$ are equal to the value of their dimensional counterparts divided by $k_p L$. In~\eqref{eq:nondim_vertical}, $\mu = {mg}/{\gamma}$ is a non-dimensional mass parameter, with ${\gamma} = T_d^2/T_g^2 =({c_d^2 /k_p^2)/( L/g}) \gg 1$. 


We consider the active force element $F_a$ generates either a contractile (pulling) or an extensile (pushing) force as according to the following state-dependent control law: if the tube foot reaches a length $l \leq l_{\rm min}$, the active force is zero and the tube foot cannot contract further, the controller requires that it extends by producing a pushing force $F_a$ following the profile in Fig.~\ref{fig:Fa} shown in solid line. Alternatively, if $l \geq l_{\rm max}$, the controller requires the tube foot to contract by producing a pulling force $F_a$ following the profile in Fig.~\ref{fig:Fa} shown in dashed line.

We rewrite Eq.~\eqref{eq:nondim_vertical} in light of this state-dependent controller: the expression for $F_a$ switches from pushing to pulling and vice-versa, depending on the state of the tube foot. 
We employ a change of variable  from $l$ to $\ell$ defined as follows
\begin{equation}
 \ell =  \begin{cases} l - l_{\rm min}, & \textrm{pushing}, \\
 l_{\rm max} -l, & \textrm{pulling}. \end{cases} 
 \label{eq:transform}
\end{equation}
The expressions for $\Phi_{\rm push}$ and $\Phi_{\rm pull}$, when expressed in terms of $\ell$ satisfy the symmetry property: $ \Phi_{\rm push}(\ell) = -\Phi_{\rm pull}(\ell) = \Phi(\ell)$, which follows directly from~\eqref{eq:phi} and~\eqref{eq:transform},
\begin{equation}
\begin{split}
\Phi(\ell) = \begin{cases} 
 \dfrac{\ell}{L-\Delta },  \ &  0 <\ell< L -\Delta,  \\[3ex]
 \dfrac{L - \ell}{\Delta },  \ & L-\Delta< \ell < L, \\[3ex]
 \quad 0,   & \ell < 0$ \ \textrm{and} \  $\ell > L.
\end{cases} 
\end{split}
\end{equation}
Here, $\Delta$ denotes the change in length from where the active force is maximum to where it decays to zero; see Fig.~\ref{fig:Fa}. Namely, $\Delta= l_{\rm max} - l_c$ when pushing and  $\Delta= l_c - l_{\rm min}$ when pulling, and by the symmetry property considered here, both values are equal. 
We also introduce $\delta = l_o-l_{\rm min}$ for pushing and $\delta = l_{\rm max} - l_o$ for pulling, which  we take to be equal. 
We get a simplified expression of Eq.~\eqref{eq:nondim_vertical} during pushing and pulling,
\begin{equation}
\mu \ddot{{\ell}} + c_d \dot{{\ell}} +  k_p \ell
  =
F_{\rm max} \Phi(\ell) + k_p\delta  \mp {mg}  , 
\label{eq:ext_cont}
\end{equation}
Here, $-mg$ is for pushing and $+ mg$ is for pulling. 

Eq.~\eqref{eq:ext_cont} has several important consequences. The most important is that the weight acting on the tube foot breaks the extensile/contractile symmetry of the actuator: when standing on a horizontal flat surface, gravity aids the tube foot during contraction and acts against it during extension.  Active pushing forces are imperative to carry the sea star weight but tube feet can be made to contract passively under the gravity. Indeed, experimental observations suggest that sea stars relax from actively pulling by allowing their tube feet to buckle passively under weight. When pushing and pulling are both active as in the model considered here, a weight-carrying tube foot takes a longer time to fully extend from $l_{\rm min}$ to $l_{\rm max}$ than to fully contract from $l_{\rm max}$ to $l_{\rm min}$.
Lastly, the vertical oscillations afforded by Eq.~\eqref{eq:ext_cont} are unstable to all non-vertical perturbations unless multiple tube feet are put to work together as shown later.

\subsection{Body Mechanics}

We model the sea star as a rigid body of mass $m$ connected to a series of $N$ tube feet separated by a constant distance $d$, as shown in Fig.~\ref{fig:seastar}(g). Let $(x,y)$ denote the position of the center of mass of the sea star in inertial frame $(\mathbf{e}_x,\mathbf{e}_y)$, and $\beta$ denote its tilting angle measured from the $x$-axis in the counter-clockwise direction.
The signed position of the base point of each tube foot $n$ relative to the sea star center of mass is $d_n$, such that $d_{n+1} - d_n = d$, $n=1,\ldots,N$.
The kinematic state of each tube foot is described by its length $l_n$ and inclination angle $\theta_n$ measured from the $y$-axis in the counter-clockwise direction.

The balance laws for the forces and moments acting on the sea star body are given by
\begin{equation}
\begin{split}
\label{eq:ss}
x\textrm{-dir:} \  & -c_x \dot{x} + \sum_n F_n \sin\theta_n =  \mu \ddot{x},\\ 
y\textrm{-dir:} \  & -c_y\dot{y} -  mg + \sum_n F_n \cos\theta_n = \mu \ddot{y},  \\ 
\textrm{tilt:} \  & -c_\beta \dot{\beta} + \sum_n F_n d_n \cos(\theta_n -\beta)  =  I \ddot{\beta}. 
\end{split}
\end{equation}
where  $I$ is the moment of inertia of the sea star body and $c_x$, $c_y$, and $c_\beta$ are the internal translational and rotational damping parameters, all expressed in dimensionless form. 
{Here, to simplify the problem, we don't compute the damping force $c_d \dot{l}_n$ exerted by individual tube feet. Instead we account for external damping effects from the environment in terms of lumped damping parameters $c_x$, $c_y$ and $c_\beta$.} 

The force $F_n$ exerted by tube foot $n$ on the sea star body acts along the direction of the tube foot,
\begin{equation}
\label{eq:Fn}
F_n = F_{a,n} - k_p(l_n - l_o). 
\end{equation}
The active force $F_{a,n}$ of tube foot $n$ is either a pushing or pulling force depending on its state $l_n$ and $\theta_n$. The active force profile follows directly from Eqns.~(\ref{eq:fa},\ref{eq:phi}) and it is depicted in Fig.~\ref{fig:Fa}.

To close the system of equations~\eqref{eq:ss} and~\eqref{eq:Fn}, note that the tube feet exert forces on the sea star body only when they are attached to the ground, that is to say, during the tube foot power stroke. When attached, the state $(l_n,\theta_n)$ of the tube feet must satisfy the following constraint equations
\begin{equation}
\begin{split}
\label{eq:ssc}
x_n -l_n \sin\theta_n & = x + d_n \cos\beta, \\
l_n \cos\theta_n & = y + d_n \sin\beta,
\end{split}
\end{equation}
where $x_n$ denotes the location of attachment  of tube feet $n$ on the ground.  
In this formulation, the length and orientation of the tube feet during attachment are slaved to the 
position and orientation of the sea star body.
Eqns.~\eqref{eq:ss},~\eqref{eq:Fn} and~\eqref{eq:ssc} form a differential-algebraic system of $3+2N$ equations for $3+2N$ unknowns $(x,y,\beta, l_n, \theta_n)$   provided that we define control rules for the tube feet attachment and detachment as discussed in~\S\ref{sec:control}. 


\begin{figure}[!t]
\centering
\includegraphics[scale=1]{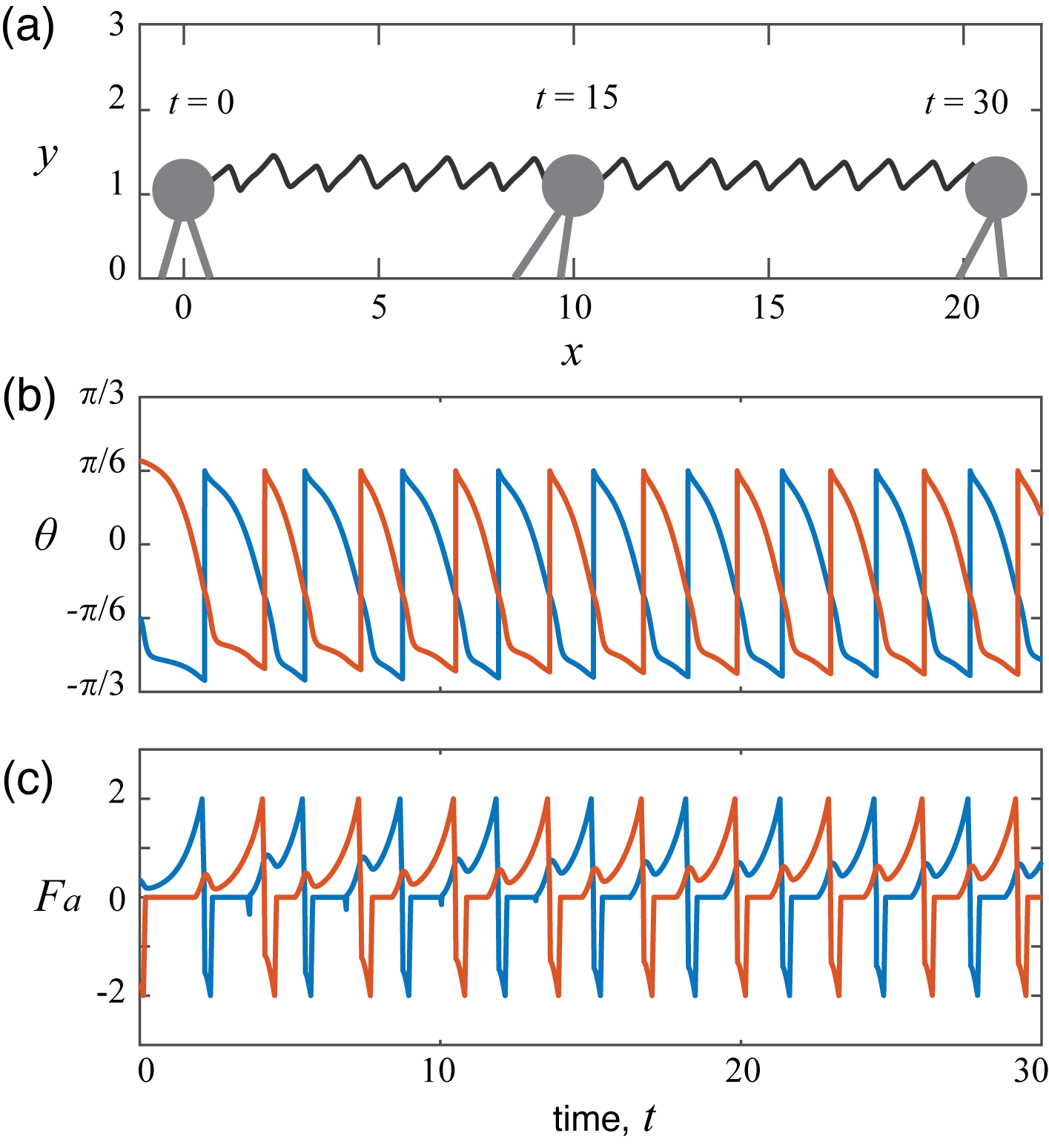}
\caption{\footnotesize \textbf{Bipedal locomotion:}  a point mass attached to two tube feet, each producing longitudinal pushing or pulling forces along the direction of the foot in the attached phase (power stroke) and taking a step forward in the detached phase (recovery stroke). (a) Trajectory of the point mass in the $(x,y)$ plane and snapshots of the walker at three instants in time. (b) Orientation angles of the tube feet versus time. (c) Active forces along the tube feet as a function of time. Positive force corresponds to pushing and negative force corresponds to pulling. The parameter values are set to $F_{\rm max} = 2$, $mg=1$ and $\gamma = 10$. The step size taken after a detachment-reattachment cycle is $\Delta \theta = \pi/6$. {(See movie S1 in the supplemental material.)} }
\label{fig:2feet}
\end{figure}


\subsection{Hierarchical Control Laws}
\label{sec:control}

We propose a hierarchical  motor control of the tube feet  consisting of {\em global} and {\em local} components: (i) a global directionality command -- descending from the nerve ring and radial nerve --  responsible for communicating the step direction to all tube feet~\citep{Smith1945}, and (ii) local sensory-motor feedback loops at the individual tube feet level that dictate the power and recovery stroke of the tube foot, that is to say, the decisions to push or pull and attach or detach. The only coupling between tube feet is via their structural attachment to the sea star body, as depicted schematically in Fig.~\ref{fig:seastar}(h). 

We implement the control law with the aforementioned global-local characteristics  into Eqns.~\eqref{eq:ss},~\eqref{eq:Fn} and~\eqref{eq:ssc} as follows. At the global sea star level, all actuators are directed using an open-loop control command that specifies the step direction $\mathbf{e}$; here the step direction is either in the negative or positive x-direction $\mathbf{e} = \pm \mathbf{e}_x$. At the local tube feet level, each actuator senses its own state $(l_n,\theta_n)$ and accordingly decides to push, pull, or detach and reattach. The local state-dependent control law can be summarized as follows. In the power stroke phase, for $l_n < l_{\rm max}$,  the  actuator $n$ decides to push or pull based on its orientation $\theta_n$ relative to the direction of motion. 
\begin{equation}
{\rm For \ } l_n < l_{\rm max}: \quad \begin{cases} \sin\theta_n \, \mathbf{e}_x\cdot \mathbf{e} > 0: \quad \textrm{pull}, \\
\sin\theta_n \, \mathbf{e}_x\cdot \mathbf{e} < 0: \quad \textrm{push} .
\end{cases}
\end{equation}
When $l_n > l_{\rm max}$, the actuator detaches, takes a step of size $\Delta \theta_n$ in the direction of motion, then reattaches to the ground (see supplemental movie S1).
These actions constitute the recovery stroke phase.
The duration of the recovery stroke, the period from detachment to reattachment, is denoted $\tau_n$. For $\tau_n = 0$, the reattachment satisfies  
\begin{equation}
x_n^+ = d_n + l_n^+ \sin\Delta \theta_n \,  \mathbf{e}_x \cdot \mathbf{e}, \qquad
l_n^+ = \dfrac{l_n^- \cos\theta_n}{\cos \Delta \theta_n}.
\end{equation}
Here, $l^-_n$ and $l^+_n$ denote the length of the tube foot right before and right after its recovery stroke, and $x_n^+$ is the point of attachment of the base of the tube foot right after recovery.

\begin{figure*}[t]
\centering
\includegraphics[scale=1]{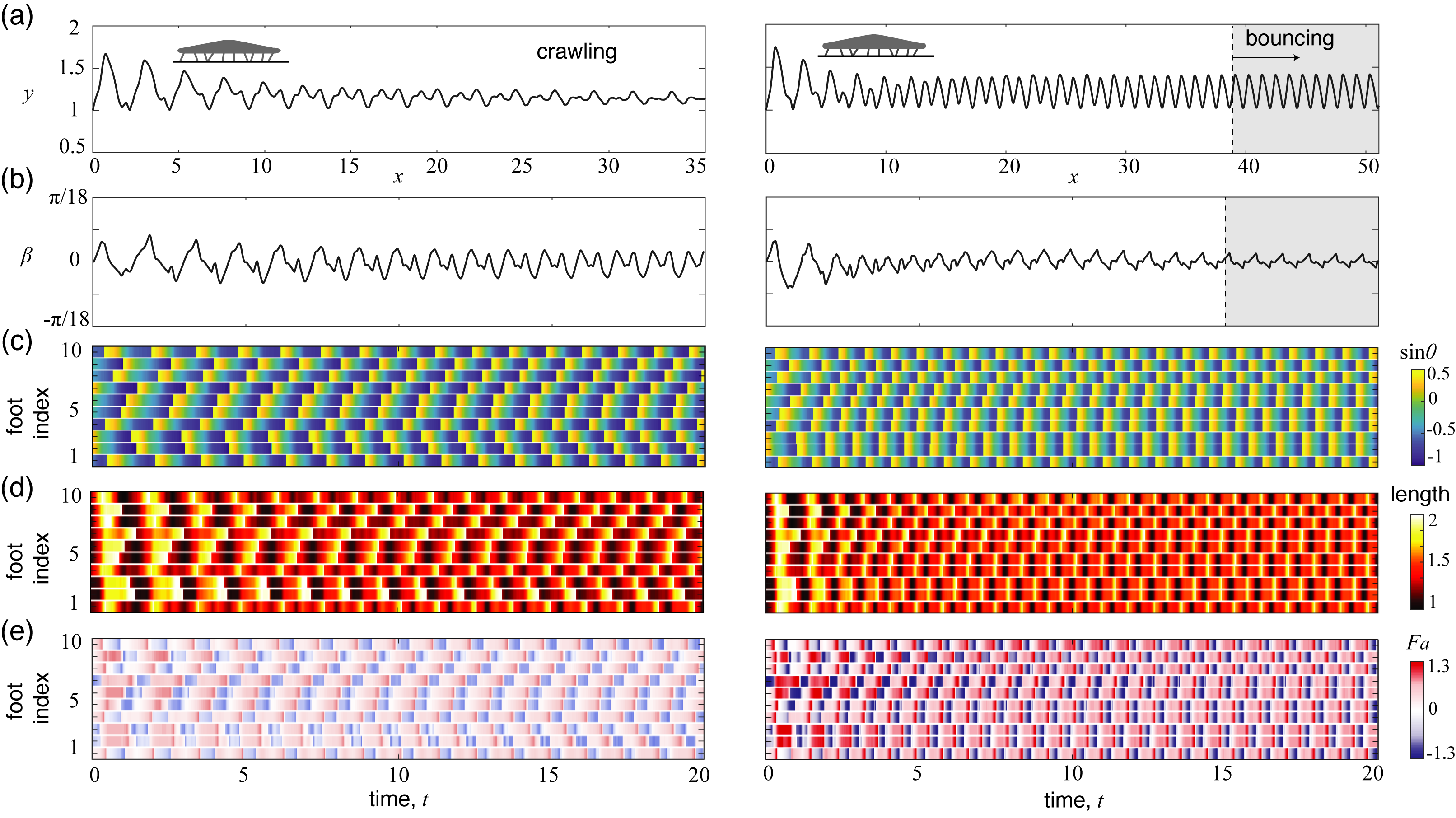}
\caption{\footnotesize \textbf{Crawl and bounce gaits:} of a sea star model with ten tube feet.  (a) Trajectory of the sea star center of mass in the $(x,y)$ plane, (b) sea star tilt angle $\beta$ versus time, (c) Tube feet orientation, (d)  Tube feet length, and (e) active forces generated along the tube feet versus time.
The active force parameter is set to $F_{\rm max} = 1$  for the results shown in the left column and $F_{\rm max} = 1.35$ in the right column; all other parameters and initial conditions are kept the same; namely, $mg=1.5$, $\gamma = 50$ and the feet are randomly oriented at $t=0$.
The sea star exhibits a crawling motion for $F_{\rm max}=1$ and it bounces for $F_{\rm max}=1.35$. 
 (See movies S2 and S3 in the supplemental material.)
}
\label{fig:crawlbounce}
\end{figure*}

\begin{figure}[t]
\centering
\includegraphics[scale=1]{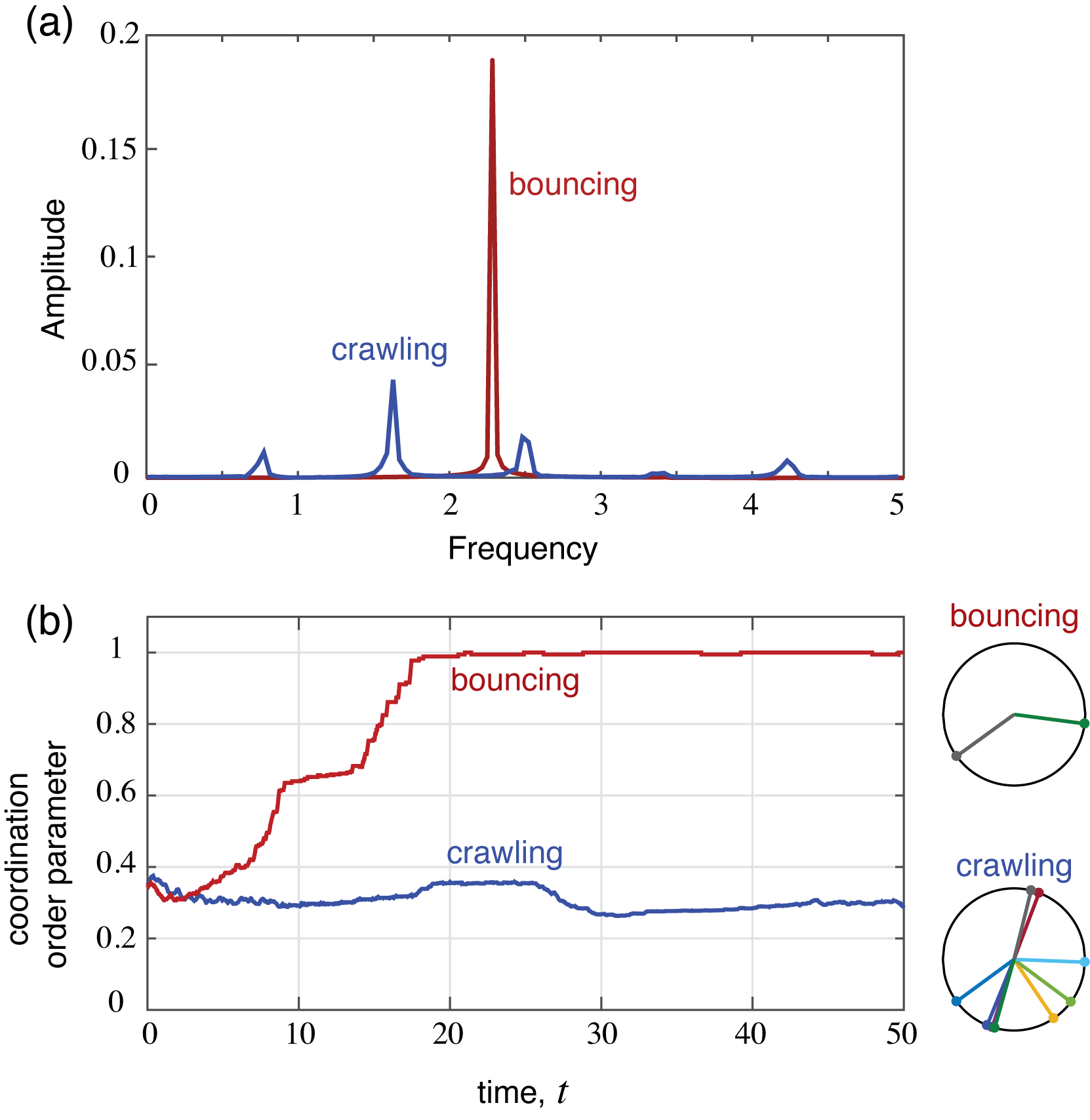}
\caption{\footnotesize \textbf{Comparison between the crawl and bounce gaits:} shown in Fig.~\ref{fig:crawlbounce}. 
(a) Frequency and amplitude of vertical oscillations, obtained by performing Fast Fourier Transform on the vertical position $y(t)$ of the sea star center of mass. In the bouncing case, vertical oscillations have a conspicuous frequency and large amplitude.
(b) Tube feet coordination order parameter versus time. Snapshot of the tube feet angles, mapped to the unit circle, are shown to the right at $t=50$. In the bouncing case, the tube feet synchronize into two clusters, which results in a high value of the coordination order parameter.
 (See movie S2 in the supplemental material.)}
\label{fig:comparison}
\end{figure}

\section{Results}
\label{sec:results}

To illustrate the hierarchical, state-dependent controller, we apply it first to the simple example of a point mass connected to two tube feet joined at their base $d=0$, as shown in Fig.~\ref{fig:2feet}(a). The two tube feet are initially oriented such that one tube foot is in a pushing state and the other in a pulling state. 
We set $F_{\rm max} = 2$, $mg = 1$, $\gamma = 10$, and $c_x = c_y = 1$. The step size $\Delta \theta = \pi/6$ is equal for both feet, and the feet have characteristic lengths $l_{\rm min} = 1$, $l_{\rm max} = 2$, $l_o = 1.5$ and $l_c = 1.9$. We follow the hierarchical control laws detailed in~\S\ref{sec:control}: both feet are instructed to step in the positive x-direction $\mathbf{e} = \mathbf{e}_x$. Other than this global directionality command, all details of the power stroke and the transition to recovery stroke  (all decisions to push or pull, or to detach and reattach) are done locally, at the tube foot level. There is no communication between the two feet other than their mechanical coupling via their attachment to the same mass. We solve the differential-algrebraic system of equations~\eqref{eq:ss} and~\eqref{eq:ssc} numerically, where the active component of $F_n$  in Eq.~\eqref{eq:Fn} is dictated by the state $(l_n,\theta_n)$ of each tube foot ($n=1,2$).
Although the controller does not explicitly impose a coordination pattern between the two feet, a clear anti-phase coordination emerged in time, and the body oscillated in the vertical direction and moved forward in the horizontal direction.
The anti-phase coordination is reflected in the angles of the tube feet and the active forces shown in Fig.~\ref{fig:2feet}(b,c). 
This walking motion is fundamentally distinct from existing models of bipedal walking \citep{Collins2001, Collins2005, Tedrake2004}: (i) the feet here are ``soft'' in the sense that they offer no resistance to bending, nor do they produce active moments during attachment; they only produce and sustain longitudinal forces along the foot length; (ii) there is no prescribed time period for attachment; the duration of each attachment cycle emerges from the state-dependent controller; (iii) the controller itself  imposes no a priori coordination between the feet. Each tube foot follows its own local sensory-motor control feedback loops, without information about the state of the other foot; coordination emerges from mechanical coupling to the point mass. We next expand on these ideas in the context of arrays of soft actuators.

We investigate the motion of the sea star model connected to ten tube feet. Specifically, we model the sea star as a rigid body, with mass $\mu$ and moment of inertia $I = 0.04 \mu D^2$, whose shape is reconstructed from a side view image of an actual sea star. The sea star damping parameters are set to $c_x = c_y = 1$, $c_\beta =10$. The tube feet are aligned in a single line, separated by distance $d= 1$, as shown in Fig.~\ref{fig:seastar}(g). The length parameters and step size of the tube feet are held at the same values as above throughout this study.
We explore the behavior of the sea star model as a function of the maximum active force $F_{\rm max}$ per tube foot,  the sea star weight $mg$, and the intrinsic damping parameter $\gamma$. 
We emphasize that the tube feet are modeled as massless actuators, that sustain and produce longitudinal forces only, with no additional constraints to prohibit intersection between neighboring feet.

{The behavior of the sea star body and tube feet is shown in Fig.~\ref{fig:crawlbounce} for $mg = 1.5$, $\gamma = 50$, $F_{\rm max} = 1$ (left column) and $F_{\rm max} = 1.35$ (right column), both starting from zero initial velocity and the same randomly-oriented feet. When $F_{\rm max} = 1$ (left column), the sea star moves in the $x$-direction, with small vertical and angular oscillations reminiscent of the crawl gait observed in actual sea stars.  For $F_{\rm max} = 1.35$ (right column) the mode of locomotion is reminiscent of the bounce gait observed in sea stars and shown in Fig.~\ref{fig:seastar}(c)~\cite{Johnson2019,Ellers2014,Ellers2018,Etzel2019}; namely, it is characterized by a distinguishable bounce frequency at the sea star level and two anti-phase clusters of tube feet, resembling the bipedal locomotion in Fig.~\ref{fig:2feet}. 
A Fast Fourier Transform of the dominant frequencies and amplitudes of vertical oscillations clearly indicate the increase in amplitude and the existence of a dominant frequency of oscillations in the bounce gait, see Fig.~\ref{fig:comparison}(a).

In crawling and bouncing, the tube feet start from the same initial orientation with no clear coordination between them in the first few steps. But, as time progresses, a coordination pattern emerges solely from the mechanical coupling between the tube feet and the sea star body. The coordination pattern is not restricted to adjacent feet, and it differs substantially between the crawling and bouncing gaits, as clearly reflected in the plots of $\sin\theta_n$, length $l_n$, and active force $F_{a,n}$ along each tube foot ($n = 1,\ldots, 10$) shown in Fig~\ref{fig:crawlbounce}(c-e). The tube feet are labeled consecutively such that two feet with labels $n$ and $n+1$ are adjacent. The feet develop a coordination pattern in time that is not restricted to adjacent feet; in the crawling motion, tube feet 2, 7 and 10 coordinate their motion while in the bouncing motion, tube feet 2, 3, 6, 7 and 9 coordinate their motion. The active forces generated in the crawling gait are weaker. The duration of the power stroke  (time from attachment to detachment) is approximately $35\%$ longer in the crawling gait than in the bouncing gait, which is consistent with our experimental observations (results not yet published). 

To quantify the degree of coordination and highlight the difference in coordination between crawling and bouncing, we sort the tube feet into subsets, or clusters, that contain tube feet of similar inclination angles $\theta_n$; namely, tube feet of angles $\theta_n$ within an angular tolerance $\epsilon=\pi/50$ from each other belong to the same cluster. The number of clusters $N_c$ lies in the range $2\leq N_c \leq N$. The case $N_c = 1$ is equivalent to a single tube foot, which cannot stably carry a weight and move forward. For $N_c = 2$, the tube feet are coordinated into two groups. For $N_c = N$, the feet exhibit maximum disorder. The degree of coordination is measured via a coordination order parameter defined as $p(t) = 2/N_c(t)$, where $p(t) \in [0.2, 1]$; $p=1$ corresponds to the tube feet split in two clusters, exhibiting the highest degree of coordination for stable locomotion (similar to bipedal locomotion), whereas lower values of $p$ indicate larger number of clusters and lower degree of coordination. 

In Fig.~\ref{fig:comparison}(b), we plot the (time-averaged) coordination order parameter $p(t)$ as a function of time for the two examples in Fig.~\ref{fig:crawlbounce}. In the bouncing gait, the coordination order parameter converges to 1 while in the crawling gait it hovers around approximately $0.3$. By way of visualization, we map the inclination angle of each tube foot to a point on the unit circle, 
$z_n(t) = e^{i \theta_n(t)}$,  for $n = 1, ..., N$,
where $z_n(t)$ indicates the position of the $n$th actuator in the complex plane. Note that the range of angles of the tube feet covers a small portion of the unit circle, since we fixed the step size to $\pi/6$. 
To make the clusters more discernible, we rescale $\theta_n$ to ${\pi}\theta_n(t)/{\theta_{\rm max}}$ to lie in the range $[0,2\pi]$. Here, $\theta_{\rm max}$ is the maximum inclination angle reached in a given simulation. A depiction of the scaled tube feet angles on the unit circle is shown for a snapshot at $t=50$ in Fig.\ref{fig:comparison}(b); clearly in the bounce gait, the tube feet angles belong to two clusters, where the feet in the same cluster are not necessarily adjacent spatially.

\begin{figure}[]
\centering
\includegraphics[scale=1]{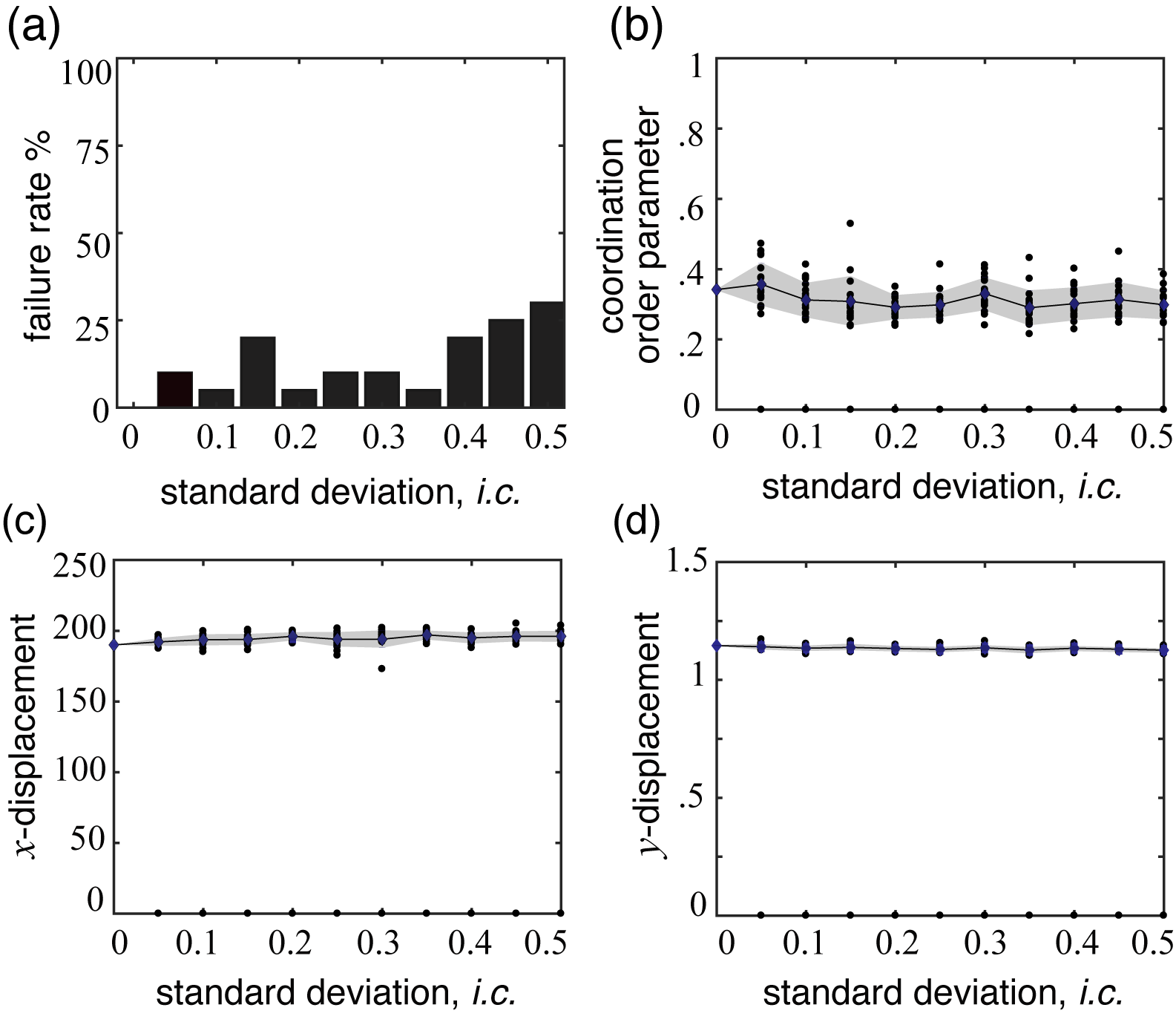}
\caption{\footnotesize  \textbf{Crawl gait: robustness to variations in tube feet initial coordination.} We randomly perturb the initial conditions of the tube feet from the case shown in Fig.~\ref{fig:crawlbounce}. The perturbations are chosen from a normal distribution with standard deviations increasing from $0$ to $0.5$ of the maximum angle $\theta_{\rm max} = \pi/3$. For each standard deviation, we perform 20 simulations, each for a total integration time of $ t = 100$. We report (a) the percentage of the initial conditions that lead to unstable motion, and for the the initial conditions that lead to stable locomotion, we report (b) the coordination order parameter, (c) the total displacement in the $x$-direction, and (d) the average vertical position. The black dots are the data points obtained from individual simulations, the line and the shaded area correspond to  the average and standard deviation of the data points, respectively.}
\label{fig:robustness_ic}
\end{figure}

\begin{figure}[]
\centering
\includegraphics[scale=1]{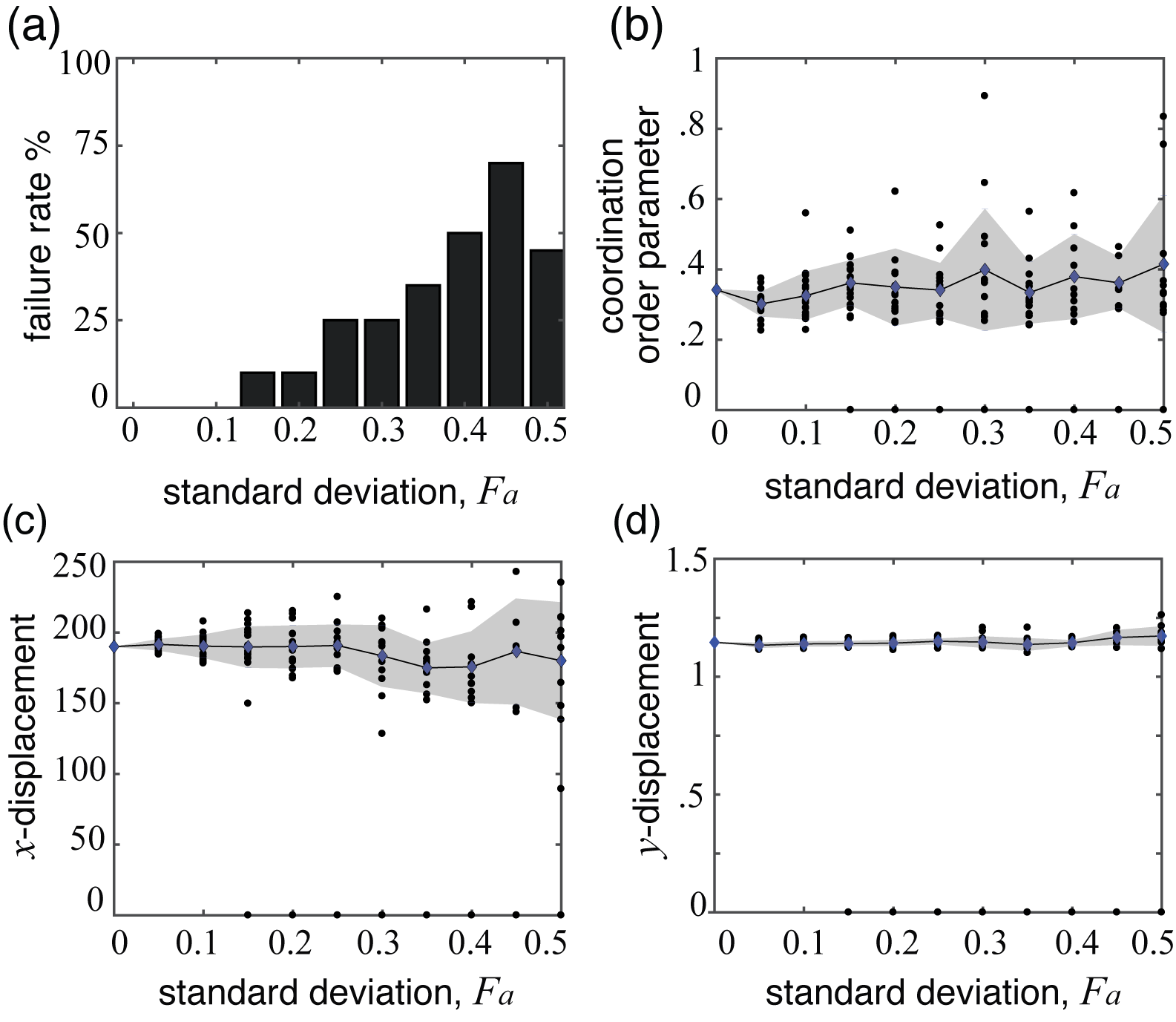}
\caption{\footnotesize  \textbf{Crawl gait: robustness to heterogeneity in the tube feet active forces.} 
We randomly perturb the maximum active force $F_{\rm max}$ in each tube foot for the crawling case shown in Fig.~\ref{fig:crawlbounce}. Each tube foot is perturbed separately, allowing for a distribution of tube feet with heterogeneous force generation ability. For each foot, $F_{\rm max}$ is chosen from a normal distribution with standard deviation equal to a fraction of  $F_{\rm max}=1$. We vary the standard deviation from $0$ to $0.5$ of $F_{\rm max}=1$. For each standard deviation, we perform 20 simulations, each for a total integration time of $t=100$. We report (a) the percentage of the initial conditions that lead to unstable motion, and for the the initial conditions that lead to stable locomotion, we report (b) the coordination order parameter, (c) the total displacement in the $x$-direction, and (d) the average vertical position. The black dots are the data points obtained from individual simulations, the line and the shaded area correspond to  the average and standard deviation of the data points, respectively.}
\label{fig:robustness_Fmax}
\end{figure}

\begin{figure*}[]
\centering
\includegraphics[scale=1]{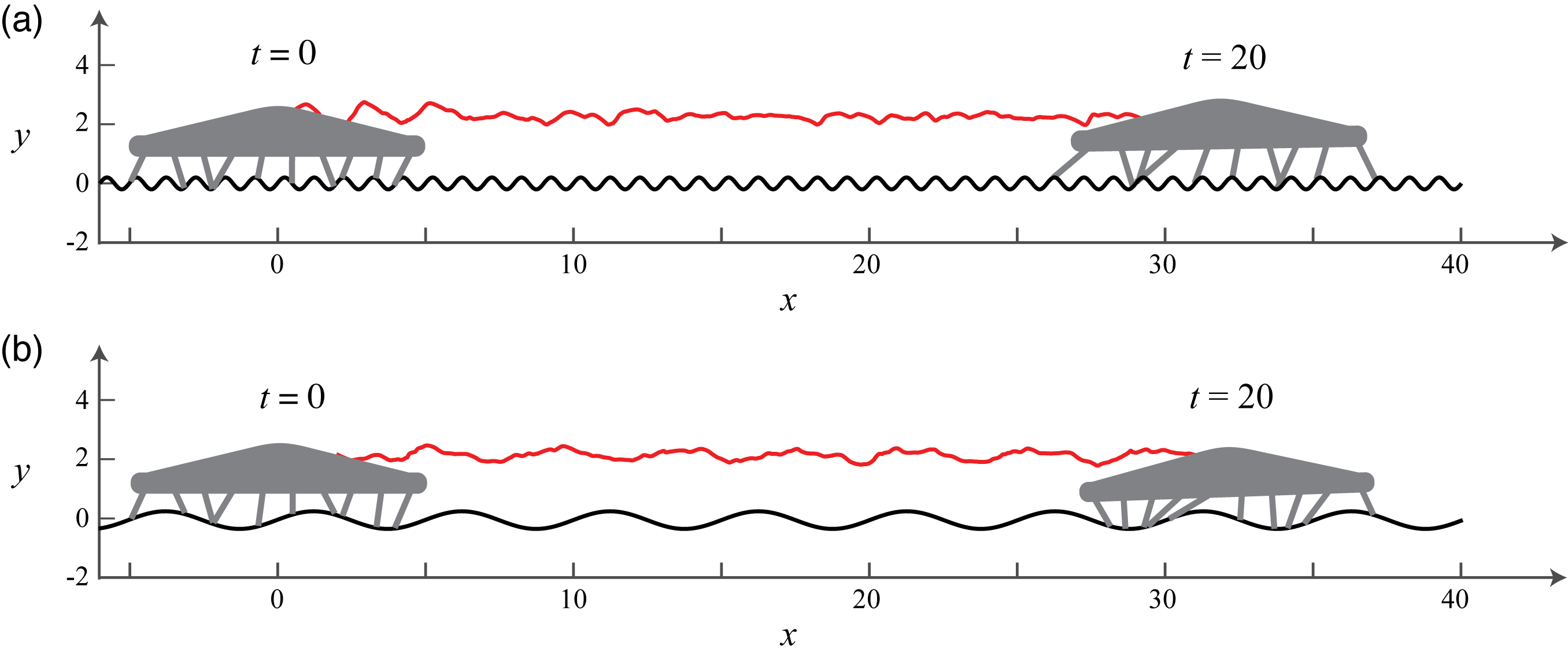}
\caption{ \textbf{Locomotion on wavy substrates}. We place the sea star model of Fig.~\ref{fig:crawlbounce} on wavy substrates without changing the parameters of the model nor the control laws. The expression for the substrates is given by (a) $y =  0.2 \sin({2 \pi x})$, and  (b) $y =  0.3 \sin({{2 \pi}x/{5}})$. {(See movie S4 in the supplemental material.)}
}
\label{fig:terrain}
\end{figure*}

\begin{figure*}[]
\centering
\includegraphics[scale=1]{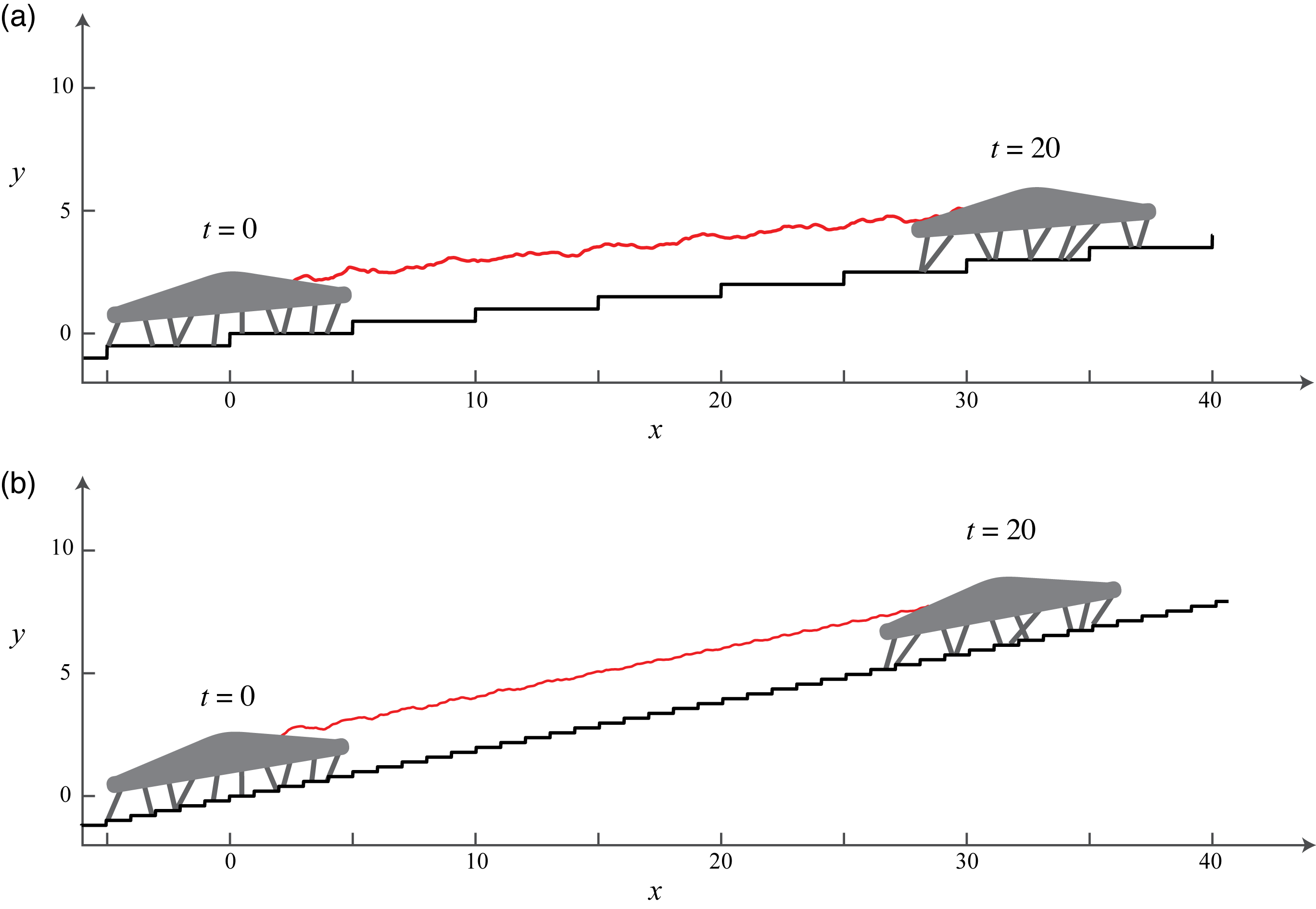}
\caption{ \textbf{Locomotion on inclined stair-like substrates}. We place the sea star model of Fig.~\ref{fig:crawlbounce} on inclined substrates without changing the parameters of the model nor the control laws.  Two types of stairs are shown:  (a) stairs whose height and width are given by height $= 0.5$ and width $= 5$, leading to an average slope of $5^\circ$, and (b) height $= 0.2$ and width $= 1$, leading to an average slope of $12^\circ$. {(See movie S5 in the supplemental material.)}
}
\label{fig:stairs}
\end{figure*}

We gauge the robustness of the crawling behavior shown in Fig.~\ref{fig:crawlbounce}(left column) to variations in the parameters of the tube feet. To this end, we perturb the initial conditions of the tube feet randomly from a normal distribution with mean values centered at the initial conditions in Fig.~\ref{fig:crawlbounce}. We vary the standard deviation from $0$ to $50 \%$ of the maximum possible initial inclination angle $\theta_{\rm max} = \pi/3$. This value of $ \theta_{\rm max}$ is set such that it automatically ensures that $l_n(0)\leq l_{\rm max}$, for all $n$.
 For each standard deviation, we perform Monte Carlo simulations with 20 random initial conditions. For a fraction of initial conditions, the sea star fails to produce stable forward movement. We report the failure rate in Fig.\ref{fig:robustness_ic}(a). The failure rate tends to increase as the standard deviation of the noise increases. For the initial conditions that produce stable locomotion, we quantify the total horizontal displacement of the body at end of the integration time, as well as the average vertical position and average coordination order parameter, both averaged over the period from $t=80$ to $t=100$. The results are shown in Fig.\ref{fig:robustness_ic}(b-d), where the black dots represent individual realizations of the Monte Carlo simulations, while the solid lines and shaded areas correspond to the mean and standard deviation of the results. It is clear from the tight standard deviations in the $x$- and $y$-displacements that the overall locomotion of the sea star is robust to variations in initial conditions, even when the details of the tube feet coordination varies.

We next explore the robustness of locomotion to heterogeneity in the tube feet actuation. Namely, we vary the active force in each tube foot independently, by choosing $F_{\rm max}$ for each tube foot randomly from a normal distribution with mean value centered at $F_{\rm max} = 1$ and a standard deviation ranging from $0$ to $50 \%$ of $F_{\rm max}$; that is, the magnitude of the active forces produced in each tube foot varies across all ten tube feet. The results of these variations on the overall sea star behavior and tube feet coordination are shown in Fig.\ref{fig:robustness_Fmax}. Similar to variations in initial condition, the failure rate generally increases with increasing standard deviation. However, in comparison to variations in initial conditions, heterogeneity in $F_a$ across tube feet produces larger variations in the tube feet coordination as well as in the overall displacement of the sea star body.

We next comment on the robustness of the crawling motion to variations in the substrate itself. We consider the sea star with the same parameters values and initial conditions shown in Fig.~\ref{fig:crawlbounce} (left panel), and we investigate its ability to crawl on wavy terrains in Fig.~\ref{fig:terrain} and up stair-like terrains in Fig.~\ref{fig:stairs}. 
The wavy substrate is described by a sinusoidal function of amplitude $a = 0.2$ and wavelength $\lambda = 1$ in Fig.~\ref{fig:terrain}(a) and $a=0.3$, $\lambda = 5$ in  Fig.~\ref{fig:terrain}(b). The stairlike terrain is described by stair width $\textrm{w} = 5$ and height $\textrm{h} = 0.5$ in Fig.~\ref{fig:stairs}(a) and $\textrm{w} = 1$, $\textrm{h} = 0.25$ in Fig.~\ref{fig:stairs}(b).  In all cases, the sea star moves robustly with adjustments made neither to the control model itself, nor to the mechanical parameters. This robustness is mediated by the decentralized local sensory-motor feedback loops at the individual tube foot level, where the control action itself depends on the state of the tube foot.

A few comments on the robustness of the bouncing gait are in order. By conducting similar numerical experiments (see supplemental movies S6-9), we found that the bouncing gait is robust for weak noise (standard deviation $\leq 10\%-15 \%$) and weak perturbations in the substrate. For larger values of noise or substrate perturbations, the distinct bouncing frequency is lost and the trajectories of stable locomotion resemble the crawling gait, albeit at the higher value of $F_{\rm max}=1.35$.  

Last, we analyze the locomotion modes on flat horizontal terrains as a function of the maximum active force $F_{\rm max}$ per tube foot, the sea star weight $mg$, and the sea star damping parameter $\gamma$.  Specifically, we look at three cross-sections of the three-dimensional parameter space $(F_{\rm max}, mg, \gamma)$, while keeping the initial conditions and all other parameter values as in Fig.~\ref{fig:crawlbounce}. In Fig.~\ref{fig:phase_diagram}(a), we investigate the sea star behavior as a function of $F_{\rm max}$ and $mg$. For weak tube feet (tube feet where $F_{\rm max}$ is small), the motion is unstable and the sea star can neither crawl nor bounce. As $F_{\rm max}$ increases for a given $mg$, the sea star first crawls, then transitions to a bouncing mode, provided that the weight exceeds a minimum value. This suggests that inertial effects, though small, seem necessary for the bouncing motion to appear. The transition from crawling to bouncing happens abruptly with the coordination order parameter increasing sharply to $1$. As $F_{\rm max}$ increases further, the motion becomes unstable again, implying that, for stable locomotion, the maximum active force per tube foot should be bounded between an upper and a lower limit. The lower limit seems to increase linearly with $mg$ for light sea stars and becomes independent of the weight as the sea star weight exceeds  $mg \approx 1.75$. Meanwhile the upper limit seems to increase linearly with $mg$, with an approximate slope of $0.7$.

The sea star behavior as a function of $F_{\rm max}$ and $\gamma$, for $mg =2$, exhibits similar trend in the transition from crawling to bouncing as $F_{\rm max}$ increases, see Fig.~\ref{fig:phase_diagram}(b). Once again, we observe that in order to achieve stable locomotion, $F_{\rm max}$ should be bounded above and below. The importance of inertial effects for bouncing is clear in these results as well. As $\gamma$ increases, inertial effects decrease, inducing a transition back to crawling for a given value of $F_{\rm max}$. 

The sea star behavior as a function of $mg$ and $\gamma$, for $F_{\rm max} =1.5$, is shown in Fig.~\ref{fig:phase_diagram}(c). The behavior is consistent with the previous observations: increasing $\gamma$ decreases the inertial effects and decreases the region of the parameter space where bouncing occurs. Further,  for a given $\gamma$, at lower load $mg$, the sea star bounces but as $mg$ increases, it transitions to crawling, similar to the effect of increasing $mg$ for a constant $F_{\rm max}$ in Fig.~\ref{fig:phase_diagram}(a).

\begin{figure*}[]
\centering
\includegraphics[scale=1]{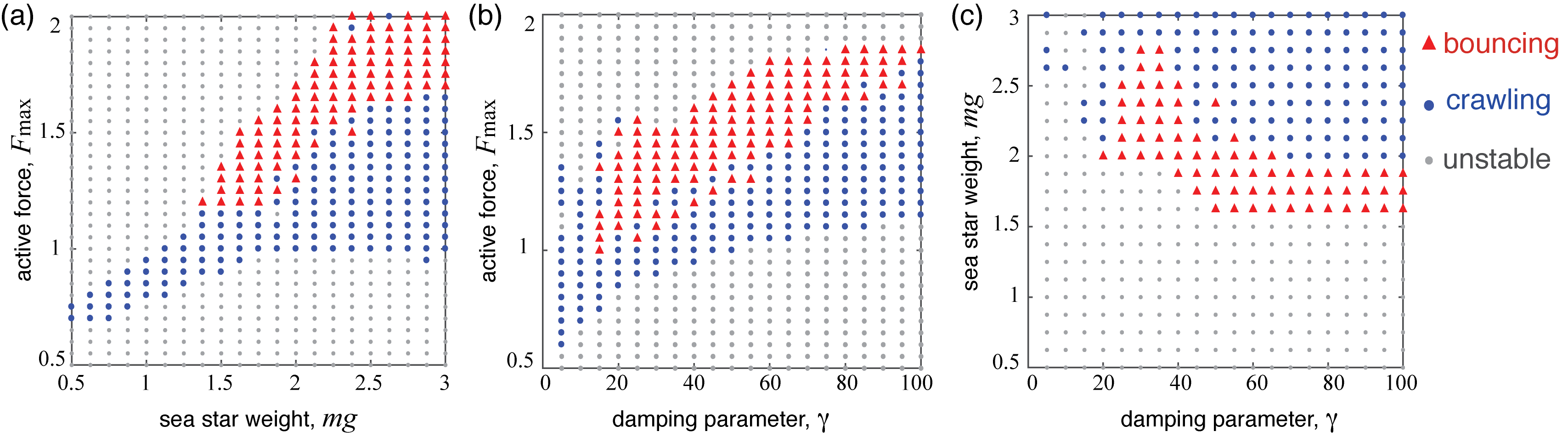}
\caption{\footnotesize \textbf{Transition from crawling to bouncing:} as a function of the maximum active force per tube foot, the sea star weight parameter spaces, and damping parameter $\gamma$ showing (a) $(mg, F_{\rm max})$ for $\gamma = 50$, (b) $(\gamma, F_{\rm max})$ for $mg = 2$, and (c) $(\gamma, mg)$ for $F_{\rm max} = 1.5$.  In all cases the initial condition is the same as in Fig.~\ref{fig:crawlbounce}.}
\label{fig:phase_diagram}
\end{figure*}

\begin{figure*}[]
\centering
\includegraphics[scale=1]{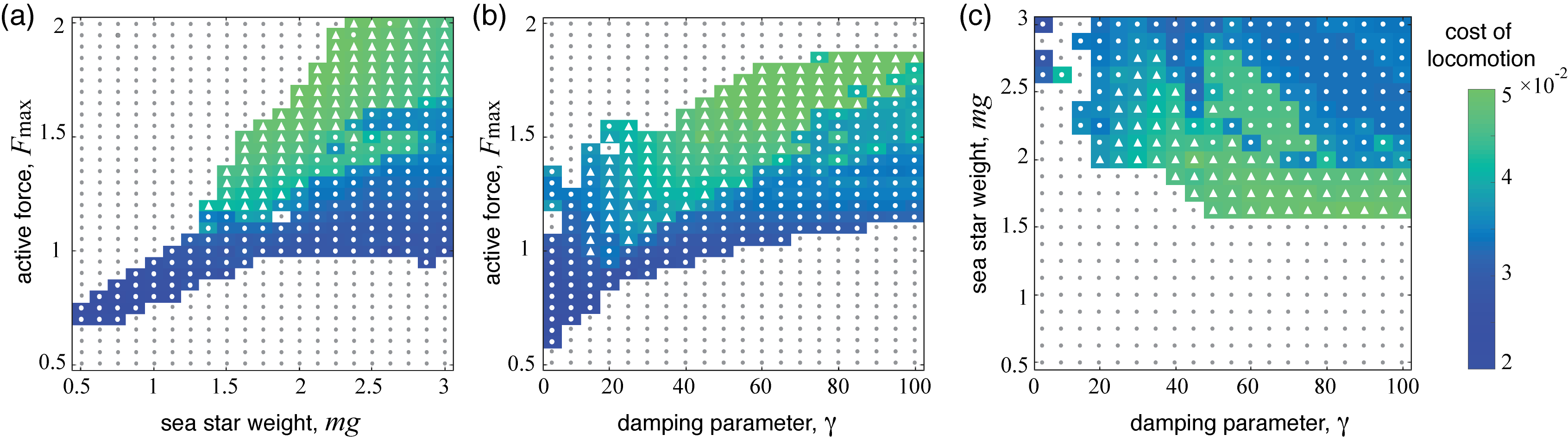}
\caption{\footnotesize \textbf{Cost of locomotion:} for the parameter spaces shown in Fig.~\ref{fig:phase_diagram}. The cost of locomotion is correlated with coordination order parameter.}
\label{fig:cost}
\end{figure*}

To examine the energetic cost of the bouncing and crawling gaits, we define cost of locomotion as the (time-averaged) active power input by all tube feet per horizontal distance traveled by the sea star, namely,
\begin{equation}
\label{eq:cost}
\textrm{cost of locomotion} = \dfrac{\langle{P_a}\rangle}{x\textrm{-distance}},
\end{equation}
where $P_a = \sum_{n} F_{a,n} \dot{l}_n$. We compute the cost of locomotion for the results in Fig.~\ref{fig:phase_diagram}, shown separately in Fig.~\ref{fig:cost} for clarity. 
The bouncing gait is correlated with a higher cost of locomotion, implying a trade-off between speed and efficiency. Bouncing gaits are characterized by higher speeds and also higher costs, which implies lower efficiency.

\section{Conclusions}
\label{sec:conc}

This study examined the control laws that underly locomotion in sea stars, as a model system for the control of distributed sensors and actuators. Sea stars use hundreds of tube feet to walk over various terrains. The tube feet seem to coordinate the direction of their power stroke, regardless of their arm's position, with the direction of walking, whereas the power and recovery strokes of individual tube feet seem to be governed locally at the tube foot level. Here, we developed a mathematical model of each tube foot as a soft actuator, consisting of active, passive, and dissipative force elements, that can actively extend or contract, generating active pulling or pushing forces on the substrate and the sea star body. We then studied the dynamics of the sea star driven by an array of such soft actuators. 
The tube feet were actuated according to a hierarchical motor control, where the direction of motion is globally communicated to all tube feet, while each foot is actuated according to local sensory-motor feedback loops. In these feedback loops, the feet use minimal sensory information (their own inclination angle and length) and generate active forces accordingly. The feet are coupled only mechanically through their structural connections to the sea star body. We found that the collective effect of the tube feet can lead to stable crawling motion of the sea star body. The model also exhibited robustness to perturbations in initial condition and heterogeneity in the ability of the tube feet to generate active forces, as well as to irregularities in the substrate geometry. 

Recent reports show that as a part of their escape response, sea stars can coordinate their numerous tube feet, in a gait known as bouncing, to increase their speed of locomotion~\cite{Johnson2019,Ellers2014,Ellers2018,Etzel2019}. We hypothesized that this transition to bouncing can occur in the context of the same hierarchical motor control used in crawling. To test this hypothesis, we varied the maximum active force $F_{\rm max}$ per tube foot, the sea star weight $mg$, and the sea star damping parameter $\gamma$. We identified a major transition in the coordination of the tube feet as we increased $F_{\rm max}$ and decreased $mg$ and $\gamma$. These transitions are invariably associated with an increase in the active work done by the tube feet relative to the work dissipated due to damping or required to lift the weight of the sea star.
During bouncing, the tube feet synchronized into two clusters, which is clearly reflected in the temporal evolution of their inclination angles, lengths, and active force. The clusters are not restricted to adjacent tube feet. Moreover, the vertical oscillations of the body were amplified, and followed a discernible frequency and wavelength; which are characteristics observed in the bounce gait in sea stars. We quantified the level of coordination in the tube feet, by introducing a coordination order parameter that takes values between 0.2 and 1. The coordination order parameter varied between 0.2 and 0.5 in the crawling motion, and stayed near $1$ in the bouncing motion.

To understand why the bounce gait is a part of the sea stars escape response as opposed to their normal mode of locomotion, we computed the cost of locomotion of the crawl and bounce gaits. We defined the cost of locomotion as the average active power consumed per horizontal distance traveled during a specific locomotion time. We found a strong correlation between the coordination order parameter and the cost of locomotion. More specifically, we found that higher tube feet coordination, characteristic of the bounce gait, consumes more power and therefore comes at a higher cost. This suggests that although  the bouncing motion can increase the speed of locomotion in sea stars, it is not always favorable for them in terms of power consumption.

A few comments on the advantages and limitations of the mathematical model are in order. Our low order model intimately couples the neural sensory-motor control to the physical system and its action on the environment, i.e, substrate. This approach is consistent with the theme of ``embodied intelligence'' or ``embodiment''~\cite{Pfeifer2006, Iida2004, Iida2007, Levy2017}. It reflects essential elements in the current understanding of how sea stars control locomotion based on neuroanatomy and behavior experiments~\cite{Smith1945,Kerkut1953, Kerkut1954, Kerkut1955} in the form of a higher level representations of the neural circuits underlying locomotion as feedback control laws.
However, our model does not describe the details of the physiology, connectivity, and activity of these neural circuits~\cite{Binyon1972, Barnes2009}. From a mechanical standpoint, our model neglects many of the complications in sea stars, including details of the tube feet biomechanics as muscular hydrostats~\cite{Kier1985, Kier1992, McHenryinprep} and deformations along the arms~\cite{Bell2018, Goldberg2019, Paschal2019, Laschi2012, Calisti2012}. Another limitation of this study is that it considers a two-dimensional model to study locomotion in one dimension. Future extensions of this work will include the more complicated dynamics required to undertake turning maneuvers.

In ongoing work, we are extracting experimental measurements from juvenile and adult sea stars in order to perform quantitative comparisons with the model. Preliminary experimental measurements support the conclusion that the bouncing gait is characterized by high values of coordination order parameter.
In addition, we are implementing a bias in the active pulling and pushing force in the model itself. This is motivated by experiments which suggest that the tube feet mostly exert pushing forces while moving on flat substrates, whereas they employ pulling forces  to walk on inclined or vertical surfaces.

We close by noting that gait transitions, reminiscent to the transition from crawling to bouncing reported here, are observed in various forms of animal locomotion including the walking to running transition in humans.  In insects, a transition from tetrapod to tripod motion is observed when walking at higher stepping frequencies. In the tripod gait, the legs coordinate into two groups:  three legs in contact with the substrate and three in a swing phase \cite{Daun-Gruhn2011, Ayali2015}. Centipedes also use numerous feet to locomote \cite{Yasui2017}, and although the underlying mechanisms for force generation {are fundamentally distinct from those of sea star tube feet}, the two systems exhibit similarities in the spatiotemporal patterns of attachment and detachment that are worth exploring in future works.

\section{Authors Contribution}

E.K. designed the research and developed the mathematical model. S.H. and E.K. conducted the research. S.H., M.M. and E.K. analyzed the results and wrote the paper. A.J. and O.E. provided panel (c) of Fig. 1 and the footage of the sea star bounce gait in Movie S3.

\section{Acknowledgements } 
E.K. and M.M. would like to acknowledge fruitful discussions with Michael Tolley, Shengqiang Cai, and Mitul Luhar.

\section{Data Accessibility}
All information needed to reproduce the results of this work are included in the main manuscript and supplemental document.

\section{Funding Statement} 
The work of S.H., M.M. and E.K. is partially supported via a Basic Research Center Grant from the Office of Naval Research, ONR Award Number: N00014-17-1-2062.

\bibliographystyle{vancouver}
\bibliography{references}

\end{document}